\documentclass[a4paper,11pt]{article}
\pdfoutput=1 

\usepackage{jcappub} 

\usepackage[T1]{fontenc} 

\let\oldhat\hat
\renewcommand{\vec}[1]{\mathbf{#1}}
\renewcommand{\hat}[1]{\oldhat{\mathbf{#1}}}

\title{$f(R)$ gravity on non-linear scales:\\
The post-Friedmann expansion and the vector potential}

\author[a,b,1]{D.B.Thomas,\note{Corresponding author.}}
\author[b]{M. Bruni,}
\author[b]{K. Koyama,}
\author[c]{B. Li,}
\author[d,b]{G.-B. Zhao}

\affiliation[a]{Department of Physics, University of Cyprus, Aglantzia, Nicosia, 2109}
\affiliation[b]{Institute of Cosmology and Gravitation, University of Portsmouth, Dennis Sciama Building, Burnaby Road, Portsmouth, PO1 3FX, UK}
\affiliation[c]{Institute for Computational Cosmology, Department of Physics, Durham University, Durham DH1 3LE, UK}
\affiliation[d]{National Astronomy Observatories, Chinese Academy of Science, Beijing, 100012, People's Republic of China}

\emailAdd{thomas.daniel@ucy.ac.cy}

\abstract{Many modified gravity theories are under consideration in cosmology as the source of the accelerated expansion of the universe and linear
perturbation theory, valid on the largest scales, has been examined in many of these models. However, smaller non-linear scales offer a richer phenomenology with
which to constrain modified gravity theories. Here, we consider the Hu-Sawicki form of $f(R)$ gravity and apply the post-Friedmann approach to derive the
leading order equations for non-linear scales, i.e. the equations valid in the Newtonian-like regime. We reproduce the standard equations for the scalar field,
gravitational slip and the modified Poisson equation in a coherent framework. In addition, we derive the equation for the leading order correction to the Newtonian
regime, the vector potential. We measure this vector potential from $f(R)$ N-body simulations at redshift zero and one, for two values of the $f_{R_0}$ parameter. We
find that the vector potential at redshift zero in $f(R)$ gravity can be close to 50\% larger than in GR on small scales for $|f_{R_0}|=1.289\times10^{-5}$, although
this is less for larger scales, earlier times and smaller values of the $f_{R_0}$ parameter. Similarly to in GR, the small amplitude of this vector potential suggests
that the Newtonian approximation is highly accurate for $f(R)$ gravity, and also that the non-linear cosmological behaviour of $f(R)$ gravity can be completely described
by just the scalar potentials and the $f(R)$ field.}

\begin{document}

\maketitle
\flushbottom

\section{Introduction}
A key problem in modern cosmology is understanding the origin of the late time accelerated expansion of the universe. One of the possible solutions to the
problem is that General Relativity (GR) is not the theory of gravity that operates on larger scales in the universe. The accelerated expansion aside,
applying GR to the entire universe is an extrapolation over many orders of magnitude compared to the scales where it is tested. For these reasons, there is
increasing interest in considering alternative gravity theories and their phenomenology, see \cite{mgreview} for a review.

On larger scales, where inhomogeneities are small and standard perturbation theory can be applied, much work has gone into showing that for many models the extra effects
can be described by a time- and space-dependent Newton's constant and ``gravitational slip'', i.e. a non-zero difference between the scalar potentials, see e.g.
\cite{ppf0,ppf1,ppf2} or further references and discussion in
\cite{mgreview,ppf3}. In addition, analytic forms can often be found for these effects. On smaller (non-linear) scales, there is potential for more phenomenology,
due to the intrinsically non-linear nature of gravity, screening mechanisms (see e.g. \cite{screening}) such as the chameleon mechanism \cite{cham,cham2} and the
possibility of significantly sourcing vector and tensor modes.

Non-linear scales are typically studied through the use of N-body simulations, either Newtonian for the case of GR, or ``Newtonian-like'' for the case of
modified gravity theories where the Newtonian gravitational equations are modified. It has been an ongoing problem in cosmology to understand how, for a GR
universe, the non-linear Newtonian gravitational equations arise, how well their solution matches that of a GR cosmology (see e.g. \cite{1407.8084}), and how
we can move beyond the Newtonian simulations to simulations that encapsulate more of GR (see e.g. \cite{1408.3352,1409.6549}). However, little attention has been given
to the same questions in modified gravity cosmologies. In addition, the derivation of the ``Newtonian-like'' equations for different modified gravity theories can be
somewhat incoherent, in the sense that several different arguments and approximations are invoked in order to simplify the equations and remove many of the terms.

One of the popular modified gravity models under consideration is $f(R)$ gravity, first suggested as a possible cause of the accelerated expansion in
\cite{carroll04,carroll05}. In this model the GR Lagrangian, consisting solely of the Ricci scalar $R$, is modified to have an additional function of the Ricci scalar
$f(R)$ included. Of particular interest is the Hu-Sawicki form for $f(R)$ gravity \cite{husawicki}, which has several nice properties: It can be tuned to match a
$\Lambda$CDM expansion history and can also evade Solar-System tests through the chameleon \cite{cham,cham2} screening mechanism. N-body simulations have been run for
the Hu-Sawicki form of $f(R)$ gravity \cite{oyaizu,1011.1257} and truly non-linear phenomena have been observed, including the manifestation of the chameleon screening
\cite{1011.1257}. As a result of this chameleon mechanism, the $f(R)$ modifications to the GR equations disappear in dense environments, which is the reason why
$f(R)$ gravity can pass Solar-System tests.

In this paper, we will use the post-Friedmann formalism \cite{postf,thesis}, a post-Newtonian type expansion in powers of the speed of light $c$, designed for a FLRW
cosmology. We will apply this formalism to $f(R)$ gravity in order to examine the Newtonian regime for this theory. In particular, we will derive the leading order
equations and the first correction to the Newtonian regime, a constraint equation for the vector potential. We will use these equations to calculate the vector
potential from N-body simulations and use this to comment on the accuracy of the ``Newtonian-like'' simulations for $f(R)$ gravity, similarly to the work carried out in
\cite{Bruni:2013mua,longerpaper} for GR.

This paper is laid out as follows. In section \ref{sec_theory} we review the pertinent details of Hu-Sawicki $f(R)$ gravity and the post-Friedmann formalism and
in section \ref{sec_analytic} we apply the post-Friedmann formalism to $f(R)$ gravity. In section \ref{sec_sims} we calculate the vector potential from $f(R)$
N-body simulations, compare it to the GR case and comment on the consequences. We conclude in section \ref{sec_conc}. Throughout the paper, a horizontal bar
denotes background quantities and $\nabla^2 A=A_{,ii}$, where the indices on the partial derivatives $_{,i}$ can be raised and lowered freely. Dots denote partial
derivatives with respect to time and a subscript $0$ denotes quantities evaluated at redshift $0$, i.e. today. Our metric sign convention is $(-+++)$.

\section{Review of relevant theory}
\label{sec_theory}
\subsection{$f(R)$ gravity}
The equations of $f(R)$ gravity are derived by generalising the Lagrangian of GR, such that a generic function of the Ricci scalar R, $f(R)$, is added to the action,
\begin{equation}
 S=\int d^4x \sqrt{-g}\frac{R+f(R)}{16\pi G}+S_{\rm{matter}}\rm{.}
\end{equation}
Specifying the function $f(R)$ then completely specifies the theory. In general, $f(R)$ gravity contains fourth order equations of motion, and is a subset of
scalar-tensor theories. The modified Einstein equations, derived by varying the new action with respect to the metric $g_{\mu \nu}$, are given by
\begin{equation}
 8\pi G T_{\mu \nu}=R_{\mu\nu}\left(1+f_R \right)-\frac{g_{\mu\nu}}{2}\left(R+f(R) \right)+g_{\mu\nu}\Box f_R-\nabla_{\mu}\nabla_{\nu}f_R \rm{,}
\end{equation}
where $\nabla_{\mu}$ is the covariant derivative, $\Box=\nabla^{\mu}\nabla_{\mu}$ and $f_R=\partial f(R)/\partial R$. $f_R$ corresponds to the extra scalar degree of
freedom that is present in $f(R)$ gravity. The Hu-Sawicki form \cite{husawicki} of $f(R)$ gravity is given by 
\begin{equation}
 f(R)=-m^2\frac{c_1\left(R/m^2 \right)^n}{c_2\left(R/m^2 \right)^n+1}\rm{.}
\end{equation}
In the limit $\vert R \vert\ \gg m^2$, this can be expanded as 
\begin{eqnarray}
&&f(R)\approx-\frac{c_1}{c_2}m^2+\frac{c_1}{c^2_2}m^2\left(\frac{m^2}{R} \right)^n\\
&&f_R=-\frac{nc_1}{c^2_2}\left(\frac{m^2}{R}\right)^{n+1} \rm{.}
\end{eqnarray}
The ratio $c_1/c_2$ is constrained by requiring that the $f(R)$ theory reproduces the $\Lambda$CDM expansion history, $m^2 c_1/c_2=16\pi G \bar{\rho}_{\Lambda}$.
The equations can be re-written in terms of $f_{R_0}$, the value of the field in the background today,
\begin{eqnarray}
&& f_R=f_{R_0}\left(\frac{\bar{R}_0}{R} \right)^{n+1}\\
&& f(R)=-16\pi G \bar{\rho}_{\Lambda}-\frac{f_{R_0}}{n}\frac{\bar{R}^{n+1}_0}{R^n}
\end{eqnarray}
Typically, the free parameter $n$ is taken to have the value unity, leaving $f_{R_0}$ as the only free parameter in the theory. As mentioned in the introduction,
This form of $f(R)$ gravity can match a $\Lambda$CDM expansion history and contains the chameleon screening mechanism \cite{cham,cham2}. In dense environments, this
mechanism screens the effects of the fifth force in the modified Einstein equations, either fully or partially. This allows the Hu-Sawicki model to fit Solar-System 
constraints \cite{husawicki}.

\subsection{Post-Friedmann formalism}
The post-Friedmann formalism was proposed in \cite{postf,thesis} and has been used to examine the vector potential in a GR$+\Lambda$CDM cosmology
\cite{Bruni:2013mua,longerpaper} and also the weak-lensing deflection angle for non-linear scales \cite{1403.4947}. It comprises a post-Newtonian type expansion of
the Einstein equations in powers of the speed of light $c$, altered compared to a Solar-System type expansion in order to apply to a FLRW cosmology. The perturbed
metric is considered in Poisson gauge and expanded up to order $c^{-5}$,
\begin{eqnarray}
 g_{00}&=&-\left[1-\frac{2U_N}{c^2}+\frac{1}{c^4}\left(2U^2_N-4U_P \right)\right]\nonumber\\
g_{0i}&=&-\frac{aB^N_i}{c^3}-\frac{aB^P_i}{c^5}\\
g_{ij}&=&a^2\left[\left(1+\frac{2V_N}{c^2}+\frac{1}{c^4}\left(2V^2_N+4V_P \right) \right)\delta_{ij} +\frac{h_{ij}}{c^4}\right]\nonumber \rm{.}
\end{eqnarray}
The two scalar potentials have each been split into their leading order (Newtonian) ($U_N$,$V_N$) and post-Friedmann ($U_P$,$V_P$) components. The vector potential
in the $0i$ part of the metric has also been split up into $B^N_i$ and $B^P_i$. As this metric is in the Poisson gauge, the three vectors $B^N_i$ and $B^P_i$ are
both divergenceless, $B^N_{i,i}=0$ and $B^P_{i,i}=0$. In addition, the tensor perturbation $h_{ij}$ is transverse and tracefree, $h^i_i=h^{,i}_{ij}=0$. From a
post-Friedmann viewpoint, the terms of order $c^{-2}$ and $c^{-3}$ are considered to be leading order and the terms of order $c^{-4}$ and $c^{-5}$ are considered to
be higher order. As is usual, the time derivatives will also come with a factor of $c^{-1}$. The formalism was designed for a $\Lambda$CDM cosmology, so the
energy-momentum tensor is constructed from the four-velocity of pressureless dust and then expanded in powers of $c$. See \cite{postf} for details of the
energy-momentum tensor, as well as useful expressions for the Ricci scalar and the other important quantities. In this paper, $\bar{\rho}$, $\delta$ and $v_i$ denote the
background density, density perturbation and velocity for the pressureless dust fluid.\\

The leading order equations ($c^{-2}$ order) derived using this formalism are equivalent to the quasi-static, weak field and low velocity limit of the Einstein (or
modified Einstein) equations. The advantage of the post-Friedmann expansion is that the corrections to this limit can be examined order by order. In a GR+$\Lambda$CDM
cosmology, the first correction is a constraint equation for the vector potential,
\begin{equation}
 \frac{1}{c^3}\nabla^2B^N_i=-\frac{16\pi G \bar{\rho} a^2}{c^3}\left[(1+\delta) v_i\right]\vert_v\rm{,}
\end{equation}
with $\vert_v$ denoting the vector (divergence-less) part. This was used to calculate the vector potential from $\Lambda$CDM N-body simulations in
\cite{Bruni:2013mua,longerpaper}. This vector potential acts as a quantitative check of the Newtonian approximation, and could also influence cosmological
observables such as weak-lensing \cite{1403.4947}.

\section{Post-Friedmann $f(R)$ gravity}\label{sec_analytic}
We will now derive the leading order gravitational equations valid for non-linear scales in $f(R)$ gravity. These equations come from expanding the modified
Einstein equations using the post-Friedmann formalism. The Ricci scalar, Ricci tensor, metric and energy-momentum tensor will all be expanded as in GR. In
addition, we need to decide how to expand the $f(R)$ and $f_R$ terms. We have some clues as to how to perform this. Since $R\sim c^{-2}$ at leading order, we
would expect $f(R)$ to be a factor of $c^{-2}$ smaller than $f_R$ at leading order. Examining the linear perturbation equations, and those currently used in
$f(R)$ N-body simulations, the $f_R$ field behaves similarly to the scalar potentials in the metric, i.e. the value of the field is small everywhere but its
spatial derivatives can be large. This suggests that at leading order, the $f_R$ field should be of similar order to the scalar potentials. Finally, considering the
form of the $f(R)$ function, we would want the leading order perturbation to occur at the same or higher order than the background term, as is the case for the Ricci
scalar and Ricci tensor. These considerations can all be satisfied by expanding the terms at leading order as follows,
\begin{eqnarray}
 &&f_R=\frac{f^*_R}{c^2}=\frac{f^*_{R_0}}{c^2}\left(\frac{\bar{R}_0}{R} \right)^2\\
&&f(R)=-\frac{16\pi G}{c^4}\bar{\rho}_{\Lambda}-\frac{f^*_{R_0}}{c^2}\bar{R}_0\frac{\bar{R}_0}{R}\rm{.}
\end{eqnarray}
In expanding the terms like this, we have assumed that the weak field approximation holds for the full $f_R$ field, both the perturbations in this field and
its background value. This holds for the cosmologically interesting Hu-Sawicki models that we are concerned with here. In addition, we have assumed that the
quasi-static approximation will hold. This has been investigated previously and found to hold on the scales that we are dealing with here \cite{li_qsa,noller_qsa}.
We now expand the modified Einstein equations to generate the leading order equations. As well as this ansatz, we note that at leading order $\bar{R}_0/R=O(1)$ and
$\bar{R}_0=O(c^{-2})$. We subtract the homogeneous background from the equations and define $\delta R^*=R^*-\bar{R}^*$ and $\delta f^*_R=f^*_R(R)-f^*_R(\bar{R})$.
Here,
$\delta R^*$ and $\delta f^*_R$ are not linearised; they are the full perturbed quantities, with only the background subtracted off.
For
the covariant derivative, we note that $\Box$ can be replaced with $\frac{1}{a^2}\nabla^2$ as the time derivative terms will be of higher order. More generally, when
two covariant derivatives are acting on a scalar $A$, $\nabla_0\nabla_0 A$ and $\nabla_{i}\nabla_{j}A$ can both be replaced by the partial derivatives, as the additional
terms that appear are higher order. However, $\nabla_0\nabla_i A=A_{,0i}-\frac{\dot{a}}{a}A_{,i}$, where now the additional term from the covariant derivative is at
the same order so must be included.
 
Then, the trace equation is
\begin{equation}
 \bar{R}\delta f_R+\delta R\left(1+f_R(\bar{R})+\delta f_R\right)-2\left(\delta R+\delta f(R)\right)+3\Box \delta f_R=\frac{8\pi G}{c^{4}} \delta T\rm{.}
\end{equation}
As we are only interested in the leading order equations here, we then expand each term as described earlier to show the leading order power of $c$ for each term
\begin{eqnarray}
&&\frac{\bar{R}^*}{c^2}\frac{\delta f^*_R}{c^2}+\frac{\delta R^*}{c^2}\left(1+\frac{f^*_R(\bar{R})}{c^2} +\frac{\delta f^*_R}{c^2}\right)-2\left(\frac{\delta R^*}{c^2}
+\frac{f^*(R)}{c^4}\right)\nonumber\\
&&+\frac{3}{a^2}\left(-\frac{\partial^2}{c^2\partial t^2}+\nabla^2\right) \frac{\delta f^*_R}{c^2}=\frac{8\pi G}{c^{4}} \left(-\bar{\rho}\delta c^2 \right)\nonumber\\
\Rightarrow&&\frac{3}{a^2}\nabla^2 \frac{\delta f^*_R}{c^2}=\frac{\delta R^*}{c^2}-\frac{8\pi G}{c^{2}}\bar{\rho} \delta\qquad\text{at leading order.}
\end{eqnarray}
We then repeat the same process for the $00$ Einstein equation,
\begin{eqnarray}
 &&\bar{R}_{00}\delta f_R+\delta R_{00}\left(1+f_R(\bar{R})+\delta f_R\right)-\frac{\bar{g}_{00}}{2}\left(\delta R+\delta f(R)\right)+
 \frac{\delta g _{00}}{2}\left(\bar{R}+\delta R+f(\bar{R})+\delta f(R)\right)\nonumber\\
 &&+\bar{g}_{00}\Box\delta f_R+\delta g _{00}\Box (f_R(\bar{R})+\delta f_R)
-\nabla_{0}\nabla_{0}\delta f_R=\frac{8\pi G}{c^{4}}\delta T_{00} \text{.}
\end{eqnarray}
Expanding out the terms and then taking only the leading order yields
\begin{eqnarray}
&&\frac{\bar{R}^*_{00}}{c^2}\frac{\delta f^*_R}{c^2}+\frac{\delta R^*_{00}}{c^2}\left(1+\frac{f^*_R(\bar{R})}{c^2}+\frac{\delta f^*_R}{c^2}\right)
+\frac{1}{2}\left(\frac{\delta R^*}{c^2}+\frac{f^*(R)}{c^4}\right)\nonumber\\
&&+\frac{U_N}{c^2}\left(\frac{\bar{R}^*}{c^2}+\frac{\delta R^*}{c^2}+\frac{f^*(\bar{R})}{c^4}
+\frac{f^*(R)}{c^4}\right)+\frac{1}{a^2}\left(\frac{\partial^2}{c^2\partial t^2}-\nabla^2\right)\frac{\delta f^*_R}{c^2}\nonumber\\
&&+\frac{2U_N}{a^2c^2}\left(-\frac{\partial^2}{c^2\partial t^2}+
\nabla^2\right)\left(\frac{f^*_R(\bar{R})}{c^2}+\frac{\delta f^*_R}{c^2}\right)-\frac{\partial}{c\partial t}\frac{\partial}{c\partial t}\frac{\delta f^*_R}{c^2}=
\frac{8\pi G}{c^{4}}\left[\bar{\rho} \delta c^2+\bar{\rho} (1+\delta) \left(v^2-2U_N \right)\right]\nonumber\\
&&\Rightarrow -\frac{1}{c^2a^2}\nabla^2U_N+\frac{1}{2}\frac{\delta R^*}{c^2}-\frac{1}{a^2}\nabla^2\frac{\delta f^*_R}{c^2}=\frac{8\pi G}{c^{2}}\bar{\rho} \delta\qquad
\text{at leading order.}
\end{eqnarray}
This equation can be combined with the trace equation to give
\begin{equation}\label{eqn_00_1}
 \frac{1}{6}\frac{\delta R^*}{c^2}-\frac{16\pi G}{3c^{2}} \bar{\rho} \delta=\frac{1}{c^2a^2}\nabla^2U_N \rm{,}
\end{equation}
and also
\begin{equation}\label{eq_tracerewritten}
 \frac{\nabla^2}{a^2} \frac{2\delta f^*_R}{c^2}-\frac{\delta R^*}{2c^2}=\frac{1}{c^2a^2}\nabla^2U_N \rm{,}
\end{equation}
which will we use shortly. A further form for the $f(R)$ Poisson equation is given by
\begin{equation}\label{eq_fr_poisson}
 \frac{8\pi G a^2}{c^2} \bar{\rho} \delta=\frac{1}{c^2}\nabla^2 \delta f^*_{R}-\frac{2}{c^2}\nabla^2 U_N \text{.}
\end{equation}
Finally, we need to examine the diagonal $ii$ Einstein equations in the same way (note that there is no summation here yet),
\begin{eqnarray}
 &&\bar{R}_{ii}\delta f_R+\delta R_{ii}\left(1+f_R(\bar{R})+\delta f_R\right)-\frac{\bar{g}_{ii}}{2}\left(\delta R+\delta f(R)\right)
 -\frac{\delta g_{ii}}{2}\left(\bar{R}+\delta R+f(\bar{R})+\delta f(R)\right)\nonumber\\
 &&+\bar{g}_{ii}\Box\delta f_R+\delta g_{ii}\Box (f_R(\bar{R})+\delta f_R) -\nabla_{i}\nabla_{i}\delta f_R=\frac{8\pi G}{c^{4}} \delta T_{ii}\text{.}
 \end{eqnarray}
 Again, expanding out and then keeping only the leading order terms gives
 \begin{eqnarray}
&&\hspace{-0.9cm}\frac{\bar{R}^*_{ii}}{c^2}\frac{\delta f^*_R}{c^2}+\frac{\delta R^*_{ii}}{c^2}\left(1+\frac{f^*_R(\bar{R})}{c^2}+\frac{\delta f^*_R}{c^2}\right)
-\frac{a^2}{2}\left(\frac{\delta R^*}{c^2}+\frac{f^*(R)}{c^4}\right)-\frac{a^2V_N}{c^2}\left(\frac{\bar{R}^*}{c^2}+\frac{\delta R^*}{c^2}+\frac{f^*(\bar{R})}{c^4}
+\frac{f^*(R)}{c^4}\right)\nonumber\\
&&\hspace{-0.9cm}+\left(-\frac{\partial^2}{c^2\partial t^2}+\nabla^2\right)\frac{\delta f^*_R}{c^2}
+\frac{2a^2V_N}{c^2}\left(-\frac{\partial^2}{c^2\partial t^2}+\nabla^2\right) \left(\frac{f^*_R(\bar{R})}{c^2}+\frac{\delta f^*_R}{c^2}\right)
-\nabla_{i}\nabla_{i}\frac{\delta f^*_R}{c^2}=\frac{8\pi G}{c^{4}}\bar{\rho}a^2 (1+\delta) v_i v_i\nonumber\\
&&\hspace{-0.9cm}\Rightarrow\frac{1}{c^2}\left[-\nabla^2V_N +\left(U_N-V_N \right)_{,ii}\right]-\frac{a^2}{2}\frac{\delta R^*}{c^2}+\nabla^2\frac{\delta f^*_R}{c^2}
-\nabla_{i}\nabla_{i}\frac{\delta f^*_R}{c^2}=0 \qquad \text{at leading order.}
\end{eqnarray}
We can then sum the three $ii$ equations and combine with equation \ref{eq_tracerewritten} to give
\begin{equation}
 \frac{1}{c^2}\nabla^2\left(U_N-V_N \right)=\frac{1}{c^2}\nabla^2 \delta f^*_R \rm{.}
\end{equation}
The final set of leading order equations is thus
\begin{eqnarray}\label{eqn_final}
&&\frac{\nabla^2}{a^2} \frac{\delta f^*_R}{c^2}=\frac{1}{3}\left(\frac{\delta R^*}{c^2}-\frac{8\pi G}{c^{2}} \bar{\rho} \delta\right)\nonumber\\
&&\frac{1}{6}\frac{\delta R^*}{c^2}-\frac{16\pi G}{3c^{2}}\bar{\rho} \delta=\frac{1}{c^2a^2}\nabla^2U_N\nonumber\\
 &&\frac{1}{c^2}\nabla^2\left(U_N-V_N \right)=\frac{1}{c^2}\nabla^2 \delta f^*_R\rm{.}
\end{eqnarray}
As expected, these post-Friedmann equations match those currently used in $f(R)$ N-body simulations. In particular, as $\delta R^*$
and $\delta f^*_R$ are not linearised, the full chameleon mechanism is incorporated in these equations. The effects of the chameleon mechanism have been seen in
$f(R)$ N-body simulations that implement these equations, see e.g. \cite{zhao_cham_nbody}.
Note that in deriving equations (\ref{eqn_final}), we
have only expanded out the Ricci tensor part of the Einstein tensor. The perturbed Ricci scalar has been kept in the equations and then used to substitute for
the $f_R$ term. This is why the potential $V_N$ that appears in the Poisson equation in \cite{postf}, here only shows up in the last of equations (\ref{eqn_final}) and
not in the Poisson equation. For $\delta f^*_R =0$, the last equation in (\ref{eqn_final}) implies the relation $V_N=U_N$ that is valid in the GR $\Lambda$CDM
context, and therefore our equations reduce to those in \cite{postf}, as they should.

A key purpose of the post-Friedmann approach is that it allows us to derive additional equations, such as the equation for the vector potential. We now apply the
same expansion to the $0i$ Einstein equation in order to derive this equation. The $0i$ Einstein equation in $f(R)$ gravity is given by
\begin{eqnarray}
&&\delta R_{0i}\left(1+f_R(\bar{R})+\delta f_R\right)-\frac{g_{0i}}{2}\left(\bar{R}+\delta R+f(\bar{R})+\delta f(R)\right)\nonumber\\
&&+g_{0i}\Box (f_R(\bar{R})
+\delta f_R)-\nabla_{0}\nabla_{i}\delta f_R=-\frac{8\pi G}{c^{3}}\bar{\rho} a(1+\delta) v_i\text{.}
\end{eqnarray}
As before, expanding out the terms and then keeping the leading order only leads us to
\begin{eqnarray}\label{eq_fr_zeroi}
&&\hspace{-0.9cm}-\frac{1}{2c^3a}\left(4\dot{a}U_{N,i}+4a\dot{V}_{N,i}-\nabla^2B^N_{i}\right) \left(1+\frac{f^*_R(\bar{R})}{c^2}+\frac{\delta f^*_R}{c^2}\right)
+\frac{aB^N_i}{2c^3}\left(\frac{\bar{R}^*}{c^2}+\frac{\delta R^*}{c^2}+\frac{f^*(\bar{R})}{c^4}+\frac{f^*(R)}{c^4}\right)\nonumber\\
&&\hspace{-0.9cm}-\frac{B^N_i}{ac^3}\left(-\frac{\partial^2}{c^2\partial t^2}+\nabla^2\right) \left(\frac{f^*_R(\bar{R})}{c^2}+\frac{\delta f^*_R}{c^2}\right)
-\frac{\partial}{c\partial t}\frac{\delta f^*_{R,i}}{c^2}+\frac{\dot{a}}{ac^3}\delta f^*_{R,i}=-\frac{8\pi G}{c^{3}}\bar{\rho} a(1+\delta) v_i\nonumber\\
&&\hspace{-0.9cm}\Rightarrow\frac{1}{c^3}\hspace{-0.1cm}\left(2\dot{a}U_{N,i}+2a\dot{V}_{N,i}-\nabla^2\frac{B^N_{i}}{2} 
+a\dot{\delta f}^*_{R,i}-\dot{a}\delta f^*_{R,i}\right)\hspace{-0.1cm}=\hspace{-0.1cm}\frac{8\pi G}{c^{3}}\bar{\rho} a^2(1+\delta) v_i\quad \text{at leading order.}
\end{eqnarray}
As in GR, the scalar part of this final equation is redundant: If we take the divergence of this equation then we get an equivalent equation to the time
derivative of equation (\ref{eq_fr_poisson}), using energy-momentum conservation to substitute for $\dot{\delta}$. Note that the continuity and Euler equations
for the matter are the same as for a $\Lambda$CDM cosmology, see \cite{postf} for these equations in the post-Friedmann approach.

Taking the vector (divergence free) part of the final equation in (\ref{eq_fr_zeroi}) leads to the following equation for the vector potential at leading order,
\begin{equation}\label{eq_frvec}
 \frac{1}{c^3}\nabla^2B^N_i=-\frac{16\pi G \bar{\rho} a^2}{c^3}\left[(1+\delta) v_i\right]\vert_v\text{.}
\end{equation}
This is the same as the equivalent equation in a GR$+\Lambda$CDM cosmology, which is easily understood as follows. At leading order in this weak-field regime, any
additional terms that contain the $f_R$ will be linear in the field. Since the new field is a scalar, no combination of derivatives of this field can create a
divergenceless quantity without multiplying the field by itself or a metric potential, which cannot occur at order $c^{-3}$. Thus, the modified Einstein equations
can only modify the scalar part of the $0i$ equation at leading order and not the vector part. We expect that this result could hold for all modified theories of gravity
that only contain scalar additional degrees of freedom. Nonetheless, the momentum field itself will evolve differently, so the source term for the vector will
have a different numerical value.

\section{Vector potential from $f(R)$ N-body simulations}\label{sec_sims}
Using equation (\ref{eq_frvec}), we can extract the vector potential from $f(R)$ N-body simulations. We will follow the extraction process as outlined in
\cite{Bruni:2013mua,longerpaper} for the momentum field method. This consists of using the Delauney Tesselation Field Estimator (DTFE) code \cite{dtfecode,dtfe1,dtfe2}
to extract the source term for the vector potential, the momentum field. After extracting the momentum field from the simulations, the scalar part is
subtracted in fourier space,
\begin{equation}
   [(1+\delta) \vec{v}]|_{v}=(1+\delta)\vec{v}-\vec{k}\frac{\left(\vec{k}\cdot[(1+\delta) \vec{v}] \right)}{k^2}\text{.}
\end{equation}
The power spectrum of the vector potential is then given by
\begin{equation}
  P_{{\vec B}^N}(k)=\left(\frac{16\pi G \bar{\rho} a^2 }{k^2}\right)^2 P_{[(1+\delta) \vec{v}]|_{v}}(k)\text{.}
\end{equation}
The momentum field can also be extracted (and the vector potential calculated) using a Cloud-in-Cells code \cite{cic} for comparison. As described in \cite{longerpaper},
there is also an additional way to extract the power spectrum of the vector potential using the DTFE code, which revolves around taking the curl of equation
(\ref{eq_frvec}). This method has also been applied to the $f(R)$ snapshots with reasonable agreement on small scales, but worse agreement on larger scales. The results
from these methods are detailed in appendix \ref{app_vec}.
For this work, the simulations were run with 1024$^3$ particles in a 250$h^{-1}$Mpc box. The initial conditions were created at redshift 49 using the Zel'dovich
approximation \cite{zelapprox}. The particles were then evolved to $z=0$ using the ECOSMOG code \cite{1110.1379} (see also \cite{1206.4317,1205.2698} for more
details), which was based on the RAMSES code \cite{0111367}. The particles were evolved under the influence of $f(R)$ gravity for two different values of the $f_{R_0}$
parameter, $|f_{R_0}|=1.289\times10^{-5}$ and $|f_{R_0}|=1.289\times10^{-6}$. The same initial distribution was also evolved under standard GR for comparison. The
cosmological parameters used for the simulation are as follows: $\Omega_m=0.267$, $\Omega_{\Lambda}=0.734$, $H_0=71$kms$^{-1}$Mpc$^{-1}$, $\sigma_8=0.8$ and $n_s=0.958$.
From \cite{Bruni:2013mua,longerpaper}, we know that the vector potential requires a good mass resolution for the calculation of the vector potential.
The resolution of the simulations used here is not quite in the ideal parameter range, so it is possible that there may be a small spurious contribution to the vector
potential power spectra extracted here. We aim to ameliorate this effect by examining the ratio of the value in the $f(R)$ and GR simulations. We have examined the
output of the simulations at both redshift zero and redshift one, and checked that the results from our GR simulation agree with the results obtained in
\cite{longerpaper}.

\subsection{Results}
\begin{figure}
\begin{center}
\includegraphics[width=2.75in,angle=270]{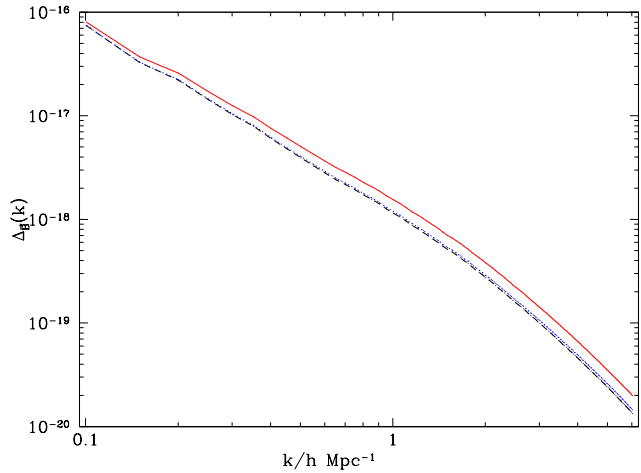}
\end{center}
\caption{The dimensionless power spectrum of the vector potential extracted from the GR and $f(R)$ N-body simulations at redshift zero. The red (solid) curve is for
$f(R)$ gravity for $|f_{R_{0}}|=1.289\times10^{-5}$ and the blue (dotted) curve is for $|f_{R_{0}}|=1.289\times10^{-6}$. The dot-dashed black curve shows the power
spectrum in GR.}
\label{fig_frvec_z0}
\end{figure}

\begin{figure}
\begin{center}
\includegraphics[width=2.75in,angle=270]{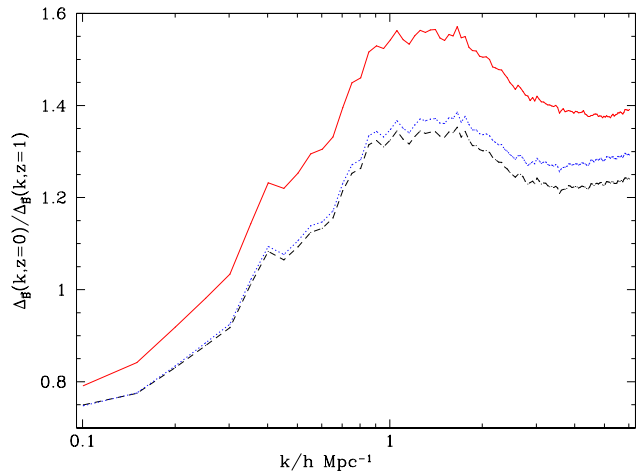}
\end{center}
\caption{The ratio of the vector potential power spectra at redshift zero and redshift one. The red (solid) curve is for $f(R)$
gravity for $|f_{R_{0}}|=1.289\times10^{-5}$ and the blue (dotted) curve is for $|f_{R_{0}}|=1.289\times10^{-6}$. The dot-dashed black curve shows the ratio for GR.}
\label{fig_frvec_z1}
\end{figure}

Our results for the vector potential are shown in figures \ref{fig_frvec_z0} and \ref{fig_frvec_z1}. The first of these shows the dimensionless vector potential power
spectrum at redshift zero, for GR (black, dashed), $|f_{R_0}|=1.289\times10^{-5}$ (red) and $|f_{R_0}|=1.289\times10^{-6}$ (blue). Figure \ref{fig_frvec_z1} shows the
ratio of the vector potentials extracted at redshift one and redshift zero for the three models. Although the vector potential is growing on non-linear scales for all 
three models, on linear scales the vector potential is actually decreasing, similarly to the linear theory scalar potential. Note that in all cases the vector potential
is much smaller than the scalar gravitational potential and shows a similar scale dependence (see \cite{longerpaper}).

\begin{figure}
\begin{center}
\includegraphics[width=2.8in,angle=270]{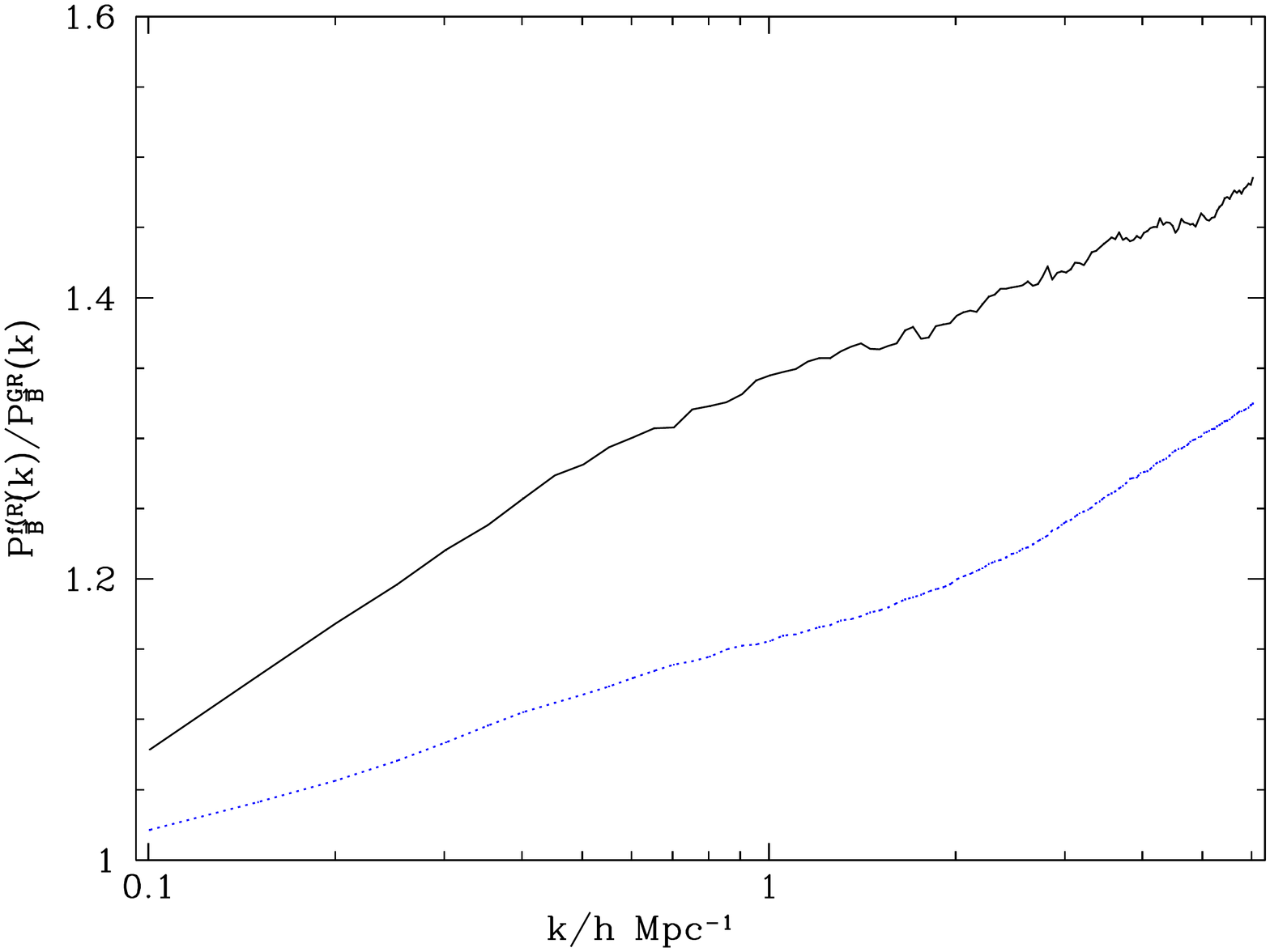}
\end{center}
\caption{The ratio of the vector potential power spectrum in $f(R)$ gravity to that in GR, for $|f_{R_{0}}|=1.289\times10^{-5}$. The blue (dotted) curve shows the
ratio at redshift one, and the black curve shows the ratio at redshift zero.}
\label{fig_frvecratio_f5}
\end{figure}

\begin{figure}
\begin{center}
\includegraphics[width=2.75in,angle=270]{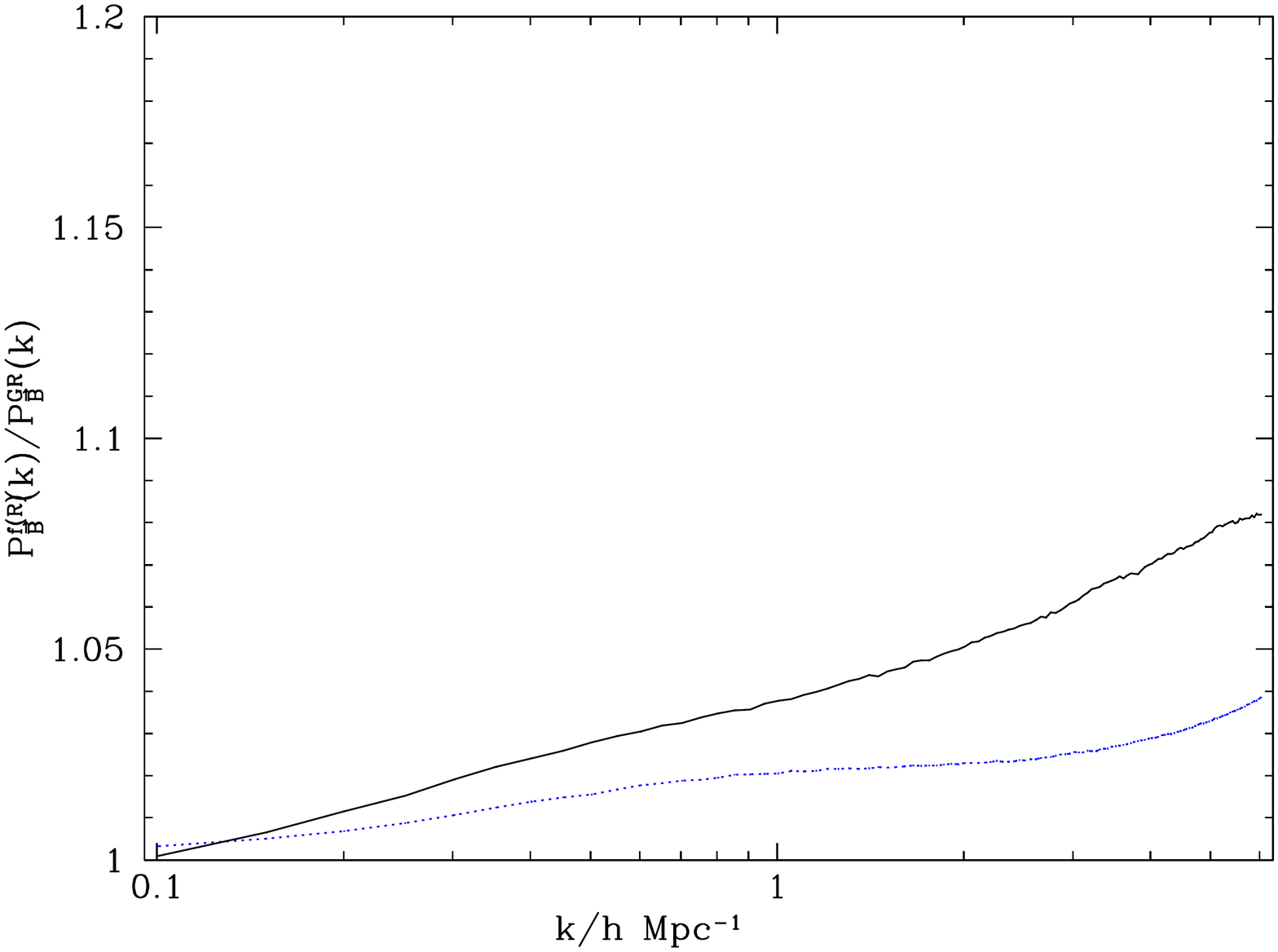}
\end{center}
\caption{The ratio of the vector potential power spectrum in $f(R)$ gravity to that in GR, for $|f_{R_{0}}|=1.289\times10^{-6}$. The blue (dotted) curve shows the
ratio at redshift one, and the black curve shows the ratio at redshift zero.}
\label{fig_frvecratio_f6}
\end{figure}

In figures \ref{fig_frvecratio_f5} and \ref{fig_frvecratio_f6} we show the ratio of the vector potential power spectrum in $f(R)$ gravity (for
$|f_{R_0}|=1.289\times10^{-5}$ and $|f_{R_0}|=1.289\times10^{-6}$ respectively) to that in GR. In these plots, the black curve shows the ratio at redshift zero and
the blue curve shows the ratio at redshift one. We can see that the vector potential power spectrum in $f(R)$ gravity is sensitive to the value of $f_{R_{0}}$, with
the difference compared to GR changing from of order a few percent to of order a few tens of percent between the two different values of $f_{R_{0}}$. In both cases,
there is a smaller difference compared to GR at earlier times and on larger scales, in line with the behaviour of the density and velocity fields in $f(R)$ gravity
\cite{1110.1379,1206.4317}.

Paralleling the GR case examined in \cite{Bruni:2013mua,1403.4947,longerpaper}, there are several immediate consequences that can be drawn from the value of the
vector potential in $f(R)$ gravity. As in GR, the vector potential is the leading order correction to the equations valid in the Newtonian-like regime, so the small
magnitude of this quantity suggests that the approximations used in deriving the equations used in $f(R)$ N-body simulations are indeed satisfied to a high degree
of accuracy. Furthermore, as noted in \cite{1403.4947}, the equation for the deflection angle in section IIIA is valid for any metric theory of gravity, including
$f(R)$ gravity. Thus, similarly to GR \cite{Bruni:2013mua,1403.4947}, the vector potential in $f(R)$ gravity is unlikely to have a significant effect on weak-lensing
observations. This also means that the standard ray-tracing approach to weak lensing should be valid in $f(R)$ gravity. Nonetheless, the larger amplitude of this
vector potential in $f(R)$ gravity on smaller scales means that, if a way can be found to observe this quantity, it can be used as an additional test of modified
gravity theories on non-linear scales.

Although little work has gone into examining $f(R)$ gravity beyond linear perturbation theory, the vector potential examined here could in principle also be calculated
using second order perturbation theory, as done for GR in \cite{0709.1619,0812.1349}. Perturbation theory would be expected to break down at most of the scales
considered here, and is wrong by nearly two orders of magnitude for GR on the smallest scales considered in \cite{Bruni:2013mua,longerpaper}. However, on the largest
scales looked at here, we are close to the linear regime, so the vector potential would be expected to agree reasonably well with the prediction from perturbation
theory. In addition, the vorticity could also be examined analytically, which is one of the contributions to the alternative curl method of extracting the vector
potential (see appendix \ref{app_vec}) and which has not yet been examined in $f(R)$ gravity. It would be interesting to perform detailed perturbative calculations in
$f(R)$ gravity in order to examine these phenomena and compare them to the results presented here. We leave this for future work.

\section{Conclusions}\label{sec_conc}
In this paper we have applied the post-Friedmann approach to $f(R)$ gravity, and in doing so have provided a coherent derivation of the equations used in $f(R)$ N-body
simulations. In additon, we have derived the equation for the first correction to the leading order equations, the vector potential. The equation for this vector
potential is the same as in GR, a result that may hold for other modified gravity theories that only contain an additional scalar degree of freedom.

We have used this equation to extract the power spectrum of the vector potential from $f(R)$ N-body simulations, following the procedure used for a GR$+\Lambda$CDM
cosmology in \cite{Bruni:2013mua,longerpaper}. The result of this analysis is shown in figures \ref{fig_frvec_z0}-\ref{fig_frvec_z1} and figures
\ref{fig_frvecratio_f5}-\ref{fig_frvecratio_f6}. We found that the vector potential can be close to 50\% larger in the $f(R)$ simulations for
$|f_{R_{0}}|=1.289\times10^{-5}$, although this is reduced at earlier times, on larger scales, or for a smaller value of $|f_{R_{0}}|$. The relatively low magnitude
of the vector potential suggests that the approximations used in deriving the equations for the N-body simulations are accurate and that the vector potential is
unlikely to be found through weak-lensing surveys.

The results of this analysis, and that of \cite{Bruni:2013mua,longerpaper}, suggests that even on scales where the density contrast is not constrained to be small,
gravitational phenomena can be entirely described by the two scalar potentials in the metric and their leading order relationship to the matter fields. This inference
should be checked for further alternative gravity theories but, if found to continue to hold, could have important consequences. In particular, although much work has
been put into model independent parameterisations of modified gravity, to date these parameterisations are perturbative and thus valid only on linear scales. If our
result holds for other theories of modified gravity, this could be used as the basis for a model independent parameterisation of modified gravity on non-linear scales.
This would be of great utility for the analysis of future surveys including the Euclid satellite.

Whilst this manuscript was being prepared, \cite{clifton_fr} appeared on the arXiv, which contains a similar $c^{-1}$ expansion applied to $f(R)$ gravity theories. The
Hu-Sawicki model studied here is in their Class I theories, for which their results seem equivalent to those obtained for the $c^{-1}$ expansion here.\\

{\sl Acknowledgements} We thank Marius Cautun for help with the publicly available DTFE code and Hector Gil Marin for provision of, and help with,
a Cloud-in-Cells code. The simulations for this paper were performed on the ICC Cosmology Machine, which is part of the DiRAC Facility jointly funded by STFC,
the Large Facilities Capital Fund of BIS, and Durham University. They were analysed on the COSMOS Shared Memory system at DAMTP, University of Cambridge
operated on behalf of the STFC DiRAC HPC Facility. This equipment is funded by BIS National E-infrastructure capital grant ST/J005673/1 and STFC grants
ST/H008586/1, ST/K00333X/1. Additionally, this work was supported by STFC grants ST/H002774/1, ST/L005573/1 and ST/K00090X/1. GBZ is supported by the 1000 Young Talents
program in China, and by the Strategic Priority Research Program “The Emergence of Cosmological Structures” of the Chinese Academy of Sciences, Grant No. XDB09000000.

\bibliographystyle{plain}
\bibliography{fR_paper_jcap}

\appendix
\section{Additional methods for extracting the vector potential}
\label{app_vec}
\begin{figure}
\begin{center}
\includegraphics[width=2.5in,angle=270]{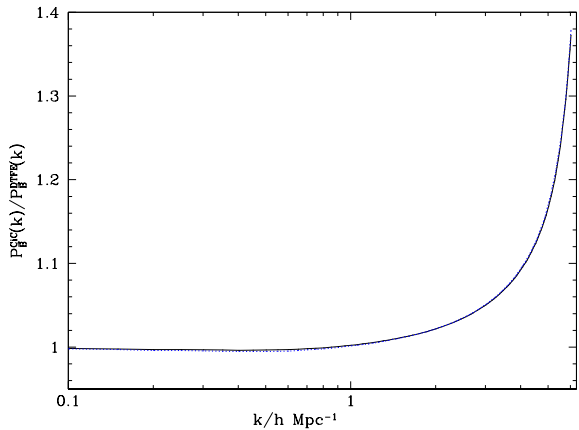}
\end{center}
\caption{The vector potential calculated from the Cloud-in-Cells momentum field divided by that calculated from the DTFE momentum field. The black curve is for GR and
the blue (dotted) curve is for $f(R)$.}
\label{fig_cicdtfe_ratio}
\end{figure}

As examined in \cite{longerpaper}, there are two methods for extracting the vector potential from N-body simulations. Here we present results from several of those
for comparison with our result in section \ref{sec_sims}. Firstly, it is possible to extract the momentum field using a Cloud-in-Cells \cite{cic} code rather than the
DTFE code. Figure \ref{fig_cicdtfe_ratio} shows the ratio between the vector potential calculated using these two methods. In this figure, the black curve shows the
ratio for GR and the blue curve shows the ratio for $f(R)$, although the two curves are almost indistinguishable. The agreement between the two methods is good for most
of the range of scales examined here, although it becomes worse for the smallest scales. This is expected due to the different window functions that
come into the two methods: The DTFE and CiC methods have a different method of smoothing the particles in the snapshot onto a regular grid, so the extracted field is
a convolution of the field with the respective window function. This begins to have a non-negligible effect on the extracted field as we move closer to the Nyquist
frequency of the grid, which is what is being seen here.

\begin{figure}
\begin{center}
\includegraphics[width=2.5in,angle=270]{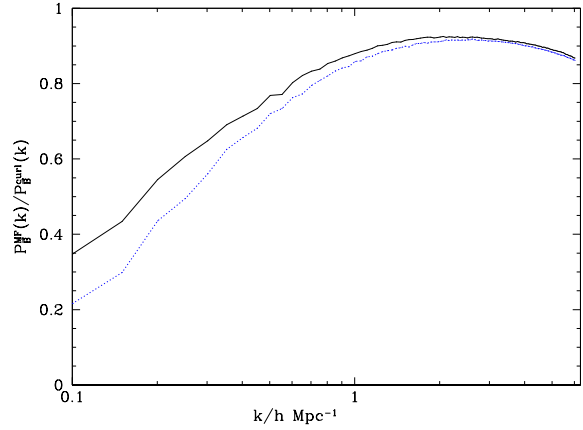}
\end{center}
\caption{The ratio of the vector potentials calculated using the momentum-field (MF) method and the curl method, for both GR (black) and $f(R)$ (blue).}
\label{fig_curlvsdtfe}
\end{figure}

In addition, rather than extracting the full momentum field and then calculating the vector part, we can take the curl of equation (\ref{eq_frvec}) as done for
GR in \cite{longerpaper}. The source term then breaks down into three components, $\nabla \delta \times \vec{v}$, $\delta \nabla \times \vec{v}$ and
$\nabla \times \vec{v}$. The advantage of splitting up the source like this for GR was two-fold: Firstly, it allowed us to compare the vorticity we are extracting with
the literature, as there were no similar results to compare to when the work of \cite{longerpaper} was carried out. Furthermore, it allowed us to show that the non-linear
quantities were contributing far more significantly than the vorticity, which only contributes linearly to the source term. Although these advantages are negated for
$f(R)$ gravity due to the lack of studies of the vorticity in the literature, we can still extract the vector potential using this method. As shown in
\cite{longerpaper}, this method does not agree exactly with the momentum field method used in the main body of the paper, with the agreement being worse on larger
scales. In figure \ref{fig_curlvsdtfe}, we show the difference between the two methods for GR (black) and $f(R)$ (blue); the agreement for GR agrees with that expected
from \cite{longerpaper}, with the $f(R)$ agreement being even worse on large scales. We are unsure why the two methods do not agree better, however we note that the
difference between the methods is insufficient to affect our conclusions regarding the validity of the approximation or the observability of the vector potential.

\begin{figure}
\begin{center}
\includegraphics[width=2.5in,angle=270]{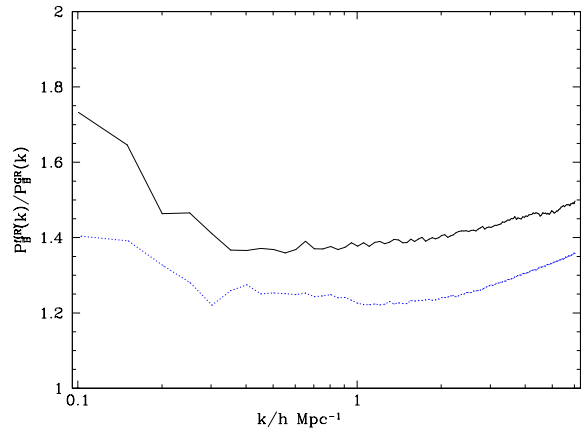}
\end{center}
\caption{The ratio between the $f(R)$ ($|f_{R_{0}}|=1.289\times10^{-5}$) and GR vector potential power spectra using the curl method. The black curve shows the ratio
at redshift zero, and the blue (dotted) curve shows the ratio at redshift one.}
\label{fig_curl_f5}
\end{figure}

In figure \ref{fig_curl_f5}, we show the analogue of figure \ref{fig_frvecratio_f5} for this ``curl'' method of extracting the vetor potential, so the black curve is
the ratio between the $f(R)$ ($|f_{R_{0}}|=1.289\times10^{-5}$) and GR power spectra at redshift $z=0$ and the blue curve is the ratio at redshift $z=1$. For most of
the range the result is similar to that obtained in the main body of the paper however, at large scales, the increased difference between the two methods for $f(R)$
gravity creates an increased ratio between the $f(R)$ and GR result. It seems unlikely that this result is physical, since $f(R)$ quantities typically return to those
of GR on these scales \cite{1110.1379,1206.4317}.

\end{document}